\documentclass[twocolumn,showpacs,preprintnumbers,amsmath,amssymb,floatfix]{revtex4}
\usepackage{graphicx}
\usepackage{wrapfig}
\input epsf

\newcommand \etal {{\it et al.\ }}
\newcommand \amp {A}
\newcommand \genop[1] {{\Lambda}^{\!#1}}
\newcommand \genopj[2] {{\Lambda}^{\!#1}_#2}
\newcommand \calnu[1] {{\cal L}_{#1}}
\newcommand \caleta[1] {{\cal L}_{#1}}

\newcommand \be {\begin{equation}}

\newcommand \ee {\end{equation}}

\newcommand \ben {\begin{eqnarray}}

\newcommand \een {\end{eqnarray}}
\newcommand \nline {\nonumber \\}

\newcommand{\beq}{\begin{equation}}
\newcommand{\eeq}{\end{equation}}

\newcommand{\beqa}{\begin{eqnarray}}
\newcommand{\eeqa}{\end{eqnarray}}

\newcommand{\comment}[1]{}

\begin{document} 

\title{Amplitude expansion of the binary phase field crystal model}
\author{K. R. Elder}
\affiliation{Department of Physics, Oakland University, 
Rochester, MI 48309}
\author{Zhi-Feng Huang}
\affiliation{Department of Physics and Astronomy, Wayne State
University, Detroit, MI 48201}
\author{Nikolas Provatas}
\affiliation{Department of Materials Science and Engineering
and Brockhouse Institute for Materials Research, McMaster
University, Hamilton, ON, Canada L8S-4L7}

\date{\today}

\begin{abstract}

Amplitude representations of a binary 
phase field crystal model are developed for a two dimensional 
triangular lattice and three dimensional BCC and FCC crystal structures.  
The relationship between these amplitude equations and the standard
phase field models for binary alloy solidification with elasticity 
are derived, providing an explicit connection between phase field 
crystal and phase field models.  Sample simulations of solute 
migration at grain boundaries, eutectic solidification and 
quantum dot formation on nano-membranes are also presented.

\end{abstract}


\maketitle

\section{Introduction}

 The development and analysis of continuum field 
theories to model complex non-equilibrium structures 
or spatial patterns has made a tremendous impact in 
many areas of research in condensed matter and 
materials physics.  A central idea in the development 
of such models is the recognition that the patterns are controlled 
by the {\it type} and {\it interaction} of {\it defects} that define 
the patterns.  For example in spinodal decomposition 
{\it surfaces} between different atomic species interact 
through {\it diffusion} in the bulk and control the phase 
segregation process.  In block copolymer 
systems \cite{block} {\it disclinations} 
interact through {\it elastic fields} and control the ordering 
of lamellar phases.  While it is desirable that  models 
of these processes be derived from some fundamental 
atomic theory, they are frequently phenomenologically 
proposed.  Classic phenomenological models include the 
Ginzburg-Landau theory of superconductivity \cite{gl50} 
and the Cahn-Hilliard-Cook theory of phase segregation \cite{ch58,c70}.  

	Several years ago a phase field model
of crystallization was phenomenologically proposed 
by exploiting the properties of free energy functionals that are 
minimized by periodic fields.  In crystallization this field is interpreted  
as the atomic number density ($\rho$), which is uniform 
in a liquid phase and is typically periodic in the solid phase.
By incorporating elasticity, dislocations and multiple crystal 
orientations, such functionals naturally incorporate the type 
and interaction of the defects that control many crystallization phenomena.
This so-called phase field crystal (PFC) model \cite{ekhg02,eg04}  
has been used to study  glass formation \cite{beg08b}, 
climb and glide dynamics \cite{bme06}, 
pre-melting at grain boundaries \cite{beg08a,mkp08}, 
epitaxial growth \cite{he08,ybv09}, commensurate/incommensurate 
transitions \cite{ak06,ramos08}, 
sliding friction phenomena \cite{ak09} and the yield 
strength of polycrystals \cite{ekhg02,eg04,htt09}.   More recently 
a simple binary phase field crystal modeled was developed \cite{epbsg07} 
that couples the features of the PFC model of a pure materials 
with a concentration 
field so that eutectic growth, spinodal decomposition and 
dendritic growth can also be studied.  This model 
can be linked with classical density functional theory (CDFT) 
(and the parameters entering CDFT), although the approximations 
needed  to connect PFC with CDFT are quite drastic.  As shown in 
recent studies on Fe \cite{jaea09} and colloidal systems 
\cite{tbvl09}, CDFT predicts 
that $\rho$ is very sharply peaked in space (at 
atomic lattice positions) while the PFC solutions 
are almost sinusoidal in space.  Nevertheless these 
same studies indicate that the parameters entering 
PFC models can be adjusted to match experimental 
quantities.

While the periodic structure of PFC models is 
essential for describing elasticity and plasticity, it is very 
inconvenient for numerical calculations.
For example PFC simulations typically require $8^d$ (where $d$ is dimension) 
spatial grid points per atomic  lattice site.  Obviously this limits 
the method to relatively small systems, although several new 
computational algorithms have been developed that can significantly 
extend the applicability of both pure \cite{brv07,cw08,wwl09,hwwl09} 
and binary PFC models \cite{rbtpfg09}.  To alleviate this limitation an
amplitude expansion of the PFC model was developed by Goldenfeld 
\etal \cite{agd05b,agd05a,agd06}.  To understand the 
idea behind such expansions it is useful to consider a one-dimensional 
equilibrium state 
of the form $n=A\sin(qx)$, where the amplitude, $A$, is zero in 
the liquid and finite in the solid state.   While the field $n$ 
varies rapidly in space, on a length scale set by $\lambda=2\pi/q$, 
the amplitude $A$ is uniform in crystalline regions and only 
varies near dislocations and liquid solid surfaces.  
Deformations of the crystal lattice can be represented by 
spatial variations in the phase of the amplitude.  Using this amplitude 
representation, Athreya \etal \cite{agdgp07} were able 
to apply adaptive mesh refinement to simulate grain growth on 
micron scales while simultaneously  resolving atomic scale 
structures at interfaces.  This remarkable achievement suggests 
that the development of amplitude expansions is very promising for 
computational materials research.  More recently this 
expansion has been extended to include spatial variations 
in the average number density in two and three 
dimensions \cite{yhet09}.

In addition to greatly increasing computational 
efficiency, amplitude representations of PFC models 
can also be exploited to establish a link between PFC type
models and traditional phase field models.  This 
link provides insight into the specific terms that enter the bulk free energy 
and gradient energy coefficients of traditional phase field models 
\cite{Wu06b,Wu07,Maj09,Pro09}.  Since the relationship between 
the parameters that enter phase field models and sharp interface models are 
well established \cite{kr98,egpk01,efkp04}, the 
relationship between parameters 
in PFC models to sharp interface models can then be established.

In this paper, amplitude expansions are developed for triangular (2d), 
BCC and FCC crystal symmetries.  The method of multiple scales expansion 
methods employed by Yeon \etal \cite{yhet09} is used. In the small deformation limit 
the expansions are shown to reproduce standard phase field 
models of solidification and eutectic growth which incorporate elasticity 
and solute segregation effects.  Sample simulations of grain boundary 
segregation, eutectic solidification and quantum dot growth 
on nano-membranes are also presented to illustrate the flexibility 
of the amplitude equations.

\section{Binary Phase Field Crystal Model}

As discussed above, the binary-alloy PFC
model developed recently \cite{epbsg07} can incorporate the important
features of solidification, phase segregation and solute expansion in
alloy systems, in addition to the elasticity, plasticity, and multiple
crystal orientations that characterize the crystalline state.
For an alloy consisting of $A$ and $B$ atoms the model can be written
in the case of equal atomic mobilities of the constituents as
\ben
\frac{\partial n}{\partial \tau} 
&=& 
M \nabla^2\frac{\delta {\cal {\cal F}}}{\delta n}
+ \zeta_n,
\nline
\frac{\partial \psi}{\partial \tau} 
&=& M \nabla^2\frac{\delta {\cal {\cal F}}}{\delta \psi}
+ \zeta_{\psi},
\label{eq:bpfc}
\een
where $M\equiv (M_A+M_B)/\rho_\ell^2$, $M_A$ ($M_B$) is 
mobility of the $A$ ($B$) species and $\tau$ is time.  The field $n$ is the 
dimensionless number density difference given by  
$n\equiv(\rho_A+\rho_B-\rho_\ell)/\rho_\ell$, 
where $\rho_A$, $\rho_B$ and $\rho_\ell$ are the atomic number densities 
of $A$ atoms, $B$ atoms and a reference liquid respectively.   
The field $\psi\approx (\rho_A-\rho_B)/\rho_\ell$, plays the role 
of a concentration field. The free energy is given by 
\ben
\frac{\Delta {\cal F}}{k_B T \rho_l} &=& \int d\vec{r} \left[
\frac{B^\ell}{2}n^2 + B^x\frac{n}{2}\Lambda n  
-\frac{t}{3}n^3
+\frac{v}{4}n^4
\right. \nline && \left.
+\gamma  \psi 
+\omega\frac{\psi^2}{2} 
+u\frac{\psi^4}{4} 
+\frac{K'}{2}|\vec{\nabla} \psi|^2
\right],
\label{eq:lazy0}
\een
where $\Lambda \equiv 2R^2\nabla^2+R^4\nabla^4$ and $t, v, \gamma, \omega, u$ 
and $K'$ are constants.  The parameters $B^\ell$,
$B^x$ and $R$ depend on $\psi$ and in the simplest non-trivial 
case can be set to $B^\ell \equiv B^\ell_0+B^\ell_2 \psi^2$, 
$B^x=B^x_0$ and  $R=R_0(1+\alpha \psi)$. Details of the 
parameters entering this model are described 
in Ref.~\cite{epbsg07}.  Here $\alpha$ is the solute 
expansion coefficient and to further simply calculations it 
will be assumed to be small.  In the small $\alpha$ limit the free energy 
can be rewritten as
\ben
F 
&=& \int d\vec{x} \left[
\frac{n}{2}\left(\genop{0}   + \alpha \psi \genop{1}\right) n  
-\frac{t}{3}n^3
+\frac{v}{4}n^4
\right. \nline  && \left. 
+\gamma  \psi 
+\frac{\omega}{2} \psi^2
+\frac{u}{4} \psi^4
+\frac{K}{2}|\vec{\nabla} \psi|^2
\right],
\label{eq:lazy1}
\een
where 
\ben
\genop{0} &\equiv& \Delta B_0 + B^\ell_2 \psi^2 + B_0^x (1+\nabla^2)^2, \nline
\genop{1} &\equiv& 4 B^x_0 (\nabla^2+\nabla^4),
\een
$\Delta B_0\equiv B^\ell_0-B^x_0$, 
$F\equiv \Delta {\cal F}/(k_B T \rho_\ell R^d_o)$, 
$\vec{x}\equiv \vec{r}/R_o$ and $K\equiv K'/R_o^2$.
The equations of motion are then,
\ben
\frac{\partial n}{\partial t} &=& \nabla^2\left[\genop{0}n
-tn^2+vn^3
+\frac{\alpha}{2}\left(\psi\genop{1}n+\genop{1}n\psi\right)\right],
\label{eq:ndyn}
\\ 
\frac{\partial \psi}{\partial t} &=& \nabla^2 \left[
 (\omega+ n^2 B^\ell_2-K\nabla^2) \psi +u \psi^3  
 + \frac{\alpha}{2} n \genop{1}n\right]
\label{eq:psidyn}
\een
where a time scale $Mk_BT\rho_\ell R_o^d$ has been adopted for the rescaling.

	In the following section the equations of motion will be 
developed for the slowly varying amplitudes that describe 
various crystalline systems. More formally, for any given 
periodic structure the density field can be expanded as
\be
n = n_0 + \sum_{\vec{G}} \eta_j e^{i\vec{G}\cdot \vec{r}} 
 + \sum_{\vec{G}} \eta^*_j e^{-i\vec{G}\cdot \vec{r}},
\ee
which separates the ``slow''-scale complex amplitudes $\eta_j$ and
average density field $n_0$ from the underlying
small-scale crystalline structure that is characterized by
 $\vec{G}=l\vec{q}_1+m\vec{q}_2+n\vec{q}_3$,
where $(\vec{q}_1,\vec{q}_2,\vec{q}_3)$ are the principle reciprocal
lattice vectors and $(l,m,n)$ are integers.  The corresponding
solutions for $n$  are then relatively smooth and can be approximated using  
only a few amplitudes.  In what follows, model equations 
will be developed for only the lowest order amplitudes that are needed to 
reconstruct a given crystal symmetry and defect structures relevant 
for controlling elastic and plastic effects in solidification and impurity segregation.  
	
	Recently many efforts have been devoted to developing amplitude 
expansion for various physical systems.  The central assumption 
of these approaches is that the amplitudes vary on scales much larger 
that the short (or ``fast'') atomic spacing scale.   Formally a small 
parameter can then be introduced that represents the ratio of the two 
scales and an expansion in this variable can be performed. 
To apply this analysis to Eqs. (\ref{eq:ndyn}) and (\ref{eq:psidyn}),
both the amplitudes ($\eta_j$) and concentration field $\psi$ are
assumed to be slow variables.  For technical details of this
multiple-scale analysis the reader can refer to Yeon \etal
\cite{yhet09} and the references therein. 
For simplicity in this paper the average atomic density (i.e., $n_0$) 
will be assumed as constant and zero, since miscibility 
gaps between liquid and solid phase can be accounted for (to 
some extent) by a miscibility in $\psi$. Moreover,  noise dynamics will be
neglected here. More complete analysis involving dynamic variation of
$n_0$, noise effects, as well as the general case of 
different atomic mobilities will be presented elsewhere \cite{hep}.

\section{Amplitude expansion for triangular symmetry}
\label{sec:amp}

	In two dimensions the free energy given in Eq. (\ref{eq:lazy1}) is 
minimized by a triangular lattice.  The corresponding principle reciprocal 
lattice vectors are given by
\ben
\vec{q}_1 = q_{eq}\left(-\hat{x} - 1/\sqrt{3}\hat{y}\right), \quad
\vec{q}_2 = q_{eq}\left(2/\sqrt{3}\hat{y}\right), 
\een
where $q_{eq} = \sqrt{3}/2$ is the equilibrium wave number.  To construct 
the minimal model of a triangular lattice only the lowest order reciprocal 
lattice vectors are needed, which correspond to $\vec{q}_1$, $\vec{q}_2$ 
and $\vec{q}_3\equiv -\vec{q}_1-\vec{q}_2=q_{eq}(\hat{x}-1/\sqrt{3}\hat{y})$.
Following the standard methods of multiple-scale expansion
\cite{re:cross93,yhet09} the following equations of motion can be derived
\ben
\frac{\partial \eta_j}{\partial t} &=& 
 {\cal L}_{j} \Bigg[\left(\genopj{0}{j} 
+3v\left(\amp^2-|\eta_j|^2\right)\right)\eta_j 
-2t\prod_{i\neq j}\eta^*_i 
 \nline && 
+\frac{\alpha}{2} \left( \psi \genopj{1}{j} \eta_j
+\genopj{1}{j}\eta_j \psi\right)
\Bigg], \nline &&
\nline
\frac{\partial  \psi}{\partial t} 
&=&  \nabla^2\Bigg[
(w+B^\ell_2 \amp^2-K\nabla^2) \psi + u  \psi^3  \nline && 
+\frac{\alpha}{2}
\sum_j\left(
\eta_j\left[\genopj{1}{j}\right]^*\eta^*_j+
\eta^*_j \genopj{1}{j} \eta_j\right) 
\Bigg],
\een
where 
\ben
\genopj{0}{j} &\equiv& \Delta B_0 + B^\ell_2\psi^2+B^{x}_0({\cal G}_j)^2  \nline
\genopj{1}{j} &\equiv& 4 B^{x}_0{\cal L}_j{\cal G}_j \nline
{\cal L}_j &\equiv& \nabla^2 + 2i\vec{q}_j\cdot\vec{\nabla} -|\vec{q}_j|^2
\nline
{\cal G}_j &\equiv& 
  \nabla^2 + 2i\vec{q}_j\cdot\vec{\nabla} \nline
\amp^2 &\equiv& 2\sum_{j=1}^{3} |\eta_j|^2
\een
To further simply calculations a long wavelength 
approximation will be made such that 
${\cal L}_j \approx -|\vec{q}_j|^2 = -1$.  Unfortunately a similar 
long wavelength approximation can't be made for ${\cal G}_j$ (e.g., 
replacing ${\cal G}_j$ by $2i\vec{q}_j\cdot\vec{\nabla}$)  
as then the equations would not be rotationally invariant.  
In this limit the equations of motion become
\ben
\frac{\partial \eta_j}{\partial t} &=& 
-\Bigg[\left(\Delta B_0+B^\ell_2\psi^2+B^x_0{\cal G}_j^2
+3v\left(\amp^2-|\eta_j|^2\right)\right)\eta_j 
\nline && 
-2t\prod_{i\neq j}\eta^*_i 
-2\alpha B^x_0\left( \psi {\cal G}_j\eta_j
+{\cal G}_j\eta_j \psi\right)
\Bigg] 
\nline
\frac{\partial  \psi}{\partial t} 
&=& \nabla^2\Bigg[
(w+B^\ell_2 \amp^2-K\nabla^2) \psi + u  \psi^3 
\nline && 
-2B^x_0 \alpha\, 
\sum_j\left(
\eta_j{\cal G}^*_j\eta^*_j+
\eta^*_j{\cal G}_j\eta_j\right) 
\Bigg].
\een
These dynamics can alternatively be written in a form of Model C type 
(in the Halperin-Hohenberg classification scheme \cite{hh77}), i.e., 
\ben
\frac{\partial \eta_j}{\partial t} 
= -\frac{\delta F}{\delta \eta_j^*}, \ \ \ \ \ \ \ \ \ 
\frac{\partial \psi}{\partial t} 
= \nabla^2 \frac{\delta F}{\delta \psi},
\label{eq:modc}
\een
where
\ben
F &=& \int d\vec{r} \Bigg[\frac{\Delta B_0}{2} \amp^2
+\frac{3v}{4}\amp^4
\nline &&
+\sum_{j=1}^{3} \left\{
B^x_0|{\cal G}_j\eta_j|^2
-\frac{3v}{2}|\eta_j|^4 \right\}
 \nline && 
 -2t\left(\prod_{j=1}^3 \eta_j+ {\rm c.c.}\right) 
+\left(\omega + B_2^\ell\amp^2\right)\frac{ \psi^2}{2} 
+\frac{u}{4} \psi^4 
 \nline && 
+ \frac{K}{2}|\vec{\nabla} \psi|^2
-2B^x_0\alpha\sum_{j=1}^3\left(
\eta_j {\cal G}^*_j\eta_j^*
+{\rm c.c.}\right) \psi \Bigg],
\label{eq:free_tri}
\een
with ${\rm c.c.}$ referring to the complex conjugate.

\subsection{Small deformation limit}
\label{sec:def}

	To gain insight into the above results and make connection
with traditional phase field models it is useful to rewrite $\eta_j$ in
the form $\eta_j= \phi\, e^{i\vec{q}_j\cdot \vec{u}}$.  In this case the 
magnitude of $\phi$ distinguishes between liquid ($\phi=0$) and 
solid phases ($\phi \neq 0$), while $\vec{u}$ is the displacement 
vector introduced in continuum elasticity theory 
\cite{ll_elasticity} to describe displacement of atoms from 
a perfectly ordered crystal lattice.
Substituting this expression into 
Eq. (\ref{eq:free_tri}) and taking the long wavelength 
limit (done by retaining derivatives only to second order) gives,
\ben
F &=& \int d\vec{r} \left[\left\{3\Delta B_0 \phi^2
-4t \phi^3 +\frac{45}{2}v\phi^4 
\right.\right.\nline && \left.\left.
+\left(\omega + 6B_2^\ell\phi^2\right)\frac{ \psi^2}{2} 
+\frac{u}{4} \psi^4 \right\}
\right.\nline && \left.
+\left\{\frac{K}{2}|\vec{\nabla}\psi|^2 
+6B^x_0|\vec{\nabla}\phi|^2\right\} 
\right.\nline && \left. 
+3B^x_0 \left\{\sum_{i=1}^2\left(
\frac{3}{2}U_{ii}^2\right)
+U_{xx}U_{yy}+2U_{xy}^2
\right\}\phi^2
\right.\nline&&\left.
+12\alpha B_0^x\left\{
-\phi\nabla^2\phi+\sum_{i=1}^2 U_{ii}\phi^2
\right\}\psi
\right],
\een
where $U_{ij}\equiv (\partial u_i/\partial x_j +
\partial u_j/\partial x_i)/2$ is the linear strain tensor. 
The first set of 
terms (defined by the $\{...\}$ brackets) is remarkably similar to 
standard phase field models developed for eutectic and dendritic 
solidification \cite{edkg94,edkg00}.  The polynomial 
in $\phi$ gives 
a first order transition from a liquid ($\phi=0$) to solid phase 
($\phi\neq 0$). 
The polynomial in $\psi$ is the typical `$\psi^4$' free energy 
used in the Cahn-Hilliard-Cook type models.  
The coupling term 
$\phi^2\psi^2$ (note: $B^\ell_2$ is negative) can lead to 
phase segregation at low temperatures when $\phi$ becomes 
large.  The second and third set of terms account for surface and 
linear elastic energy respectively. From the form of the third term it is 
straightforward \cite{ll_elasticity} 
to derive the elastic constants in dimensionless units, i.e., 
$C_{11}=C_{22}=9B^x_0\phi^2$ and $C_{12}=C_{44}=C_{11}/3$. 

	Finally the last set of terms couples the concentration 
field to the liquid/solid order parameter $\phi$ when the 
atomic species have a different size (i.e., $\alpha \neq 0$).
The term, $\psi\phi\nabla^2\phi$, 
implies preferential phase segregation to liquid/solid surface, 
grain boundaries and dislocations (i.e., regions in which 
$\phi$ varies in space).  Dynamically this term is 
responsible for solute migration at grain boundaries and solute drag.  
The last term $\psi\phi^2(U_{xx}+U_{yy})$ implies a coupling
between strain and concentration as should be expected when 
the atomic species have different sizes.  It may appear unusual 
that the free energy functional depends on the sign of the 
displacement gradients (via $U_{xx} = \partial u_x/\partial x$); 
however this sign determines whether there is a 
local compression or expansion of the lattice which would 
favor a specific atomic species based on the sign of solute expansion
coefficient $\alpha$.

\subsection{Equilibrium phase diagram}
\label{sec:equil}

\begin{figure}[btp]
\includegraphics[width=0.5\textwidth]{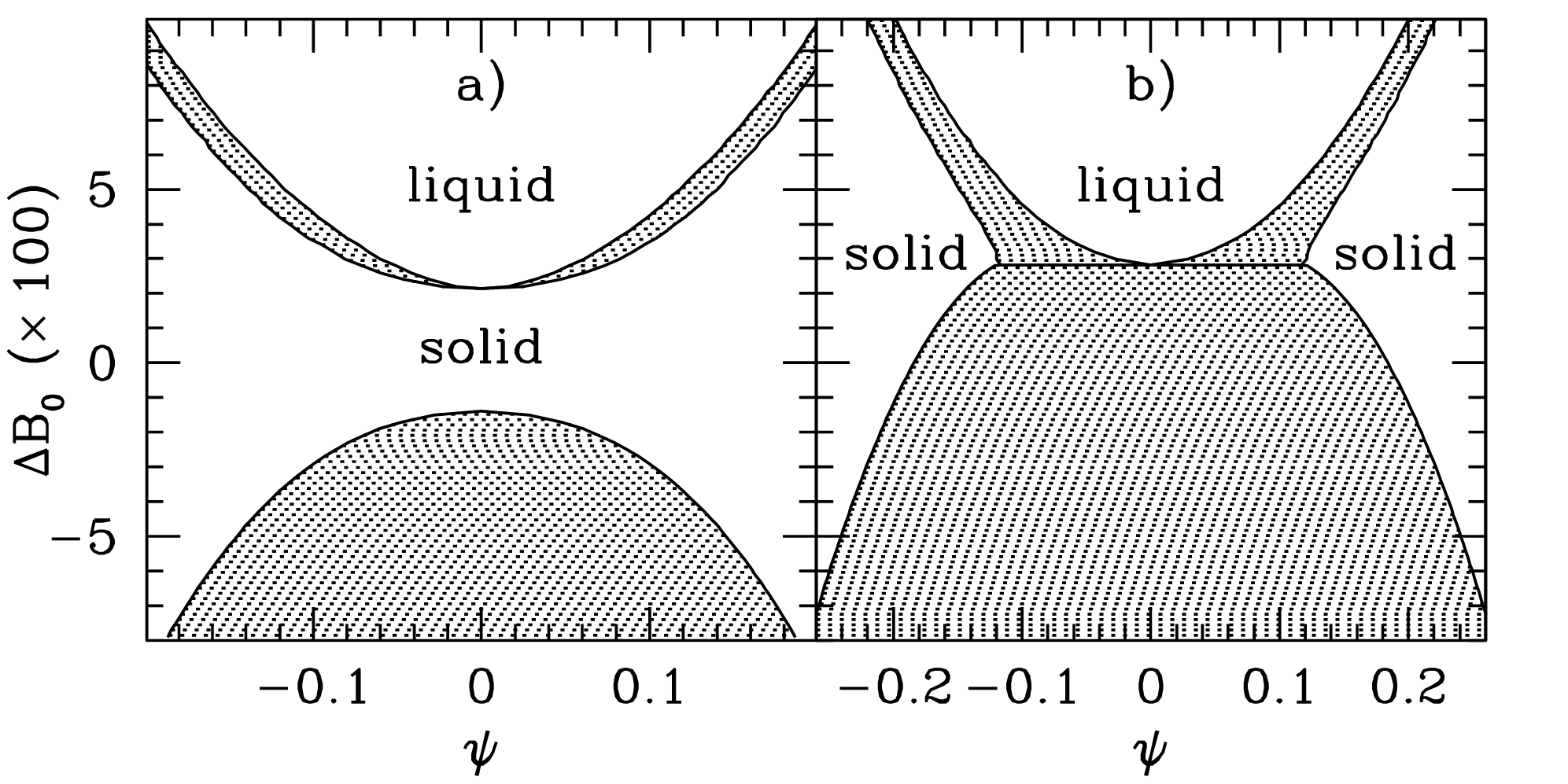}
\caption{Sample phase diagram for two dimensional 
triangular system with parameter set 
$(B^x_0,B^\ell_2,v,t,\alpha,u)=(1,-1.8,1,3/5,0.30,4)$. 
Also $\omega=0.088$ and $0.008$ in (a) and (b) 
respectively.  In each panel the filled regions correspond 
to regions of phase coexistence. }
\label{fig:triph}
\end{figure}

The equilibrium phase diagram can be evaluated by considering 
$\phi$ and $\psi$ constant and a bulk compression to 
account for solute expansion, such that $\vec{u}\equiv 
\delta (x\hat{x} +y\hat{y})$.  In this limit the free 
energy per unit area (${\cal A}$) becomes
\ben
\frac{F}{{\cal A}} &=& 3\Delta B_0 \phi^2 -4t \phi^3
+\frac{45}{2}v\phi^4 
+\frac{u}{4} \psi^4 
\nline && \!\! \!\!
+\left(\omega + 6B_2^\ell\phi^2\right)\frac{ \psi^2}{2} 
+12B^x_0\phi^2\left(\delta^2+2\alpha\psi\delta\right).\ \ 
\een
Minimizing with respect to $\delta$ gives 
$\delta_{eq} = -\alpha \psi$.  As expected the contraction/expansion 
of the lattice is controlled by $\alpha\psi$.  Substituting 
$\delta_{eq}$ for $\delta$ leads to
\ben
\frac{F}{{\cal A}} &=& 3\Delta B_0 \phi^2 -4t \phi^3
+\frac{45}{2}v\phi^4  
+\left(\omega + 6B_2^\ell\phi^2\right)\frac{ \psi^2}{2} 
\nline
&&+\frac{u}{4} \psi^4 -12B^x_0(\phi\alpha\psi)^2,
\label{eq:freet2}
\een
which is then minimized with respect to $\phi$, yielding
\be
\phi_{eq}=\frac{t+\sqrt{t^2-15v(\Delta B_0
+\psi^2(B^\ell_2-4B^x_0\alpha^2))}}{15v}.
\label{phieq}
\ee
Substituting Eq.~(\ref{phieq}) into Eq. (\ref{eq:freet2}) yields 
a free energy per unit area that is only a function of $\psi$. 
This result can be used to construct the phase diagram in 
an analogous manner as was done in Ref.~\cite{epbsg07}.
Two example phase diagrams and the corresponding model parameters
required to obtain each are show in Fig. \ref{fig:triph}.

	The derivation of amplitude equations and the related
simplifications presented above can be readily 
extended to three dimensional systems. In Appendix 
\ref{sec:BCC} and \ref{sec:FCC} the relevant 
equations and sample phase diagrams for both BCC and FCC symmetries
are presented.

\section{Applications}
\label{sec:app}

\begin{figure}[btp]
\centerline{\includegraphics[width=0.48\textwidth]{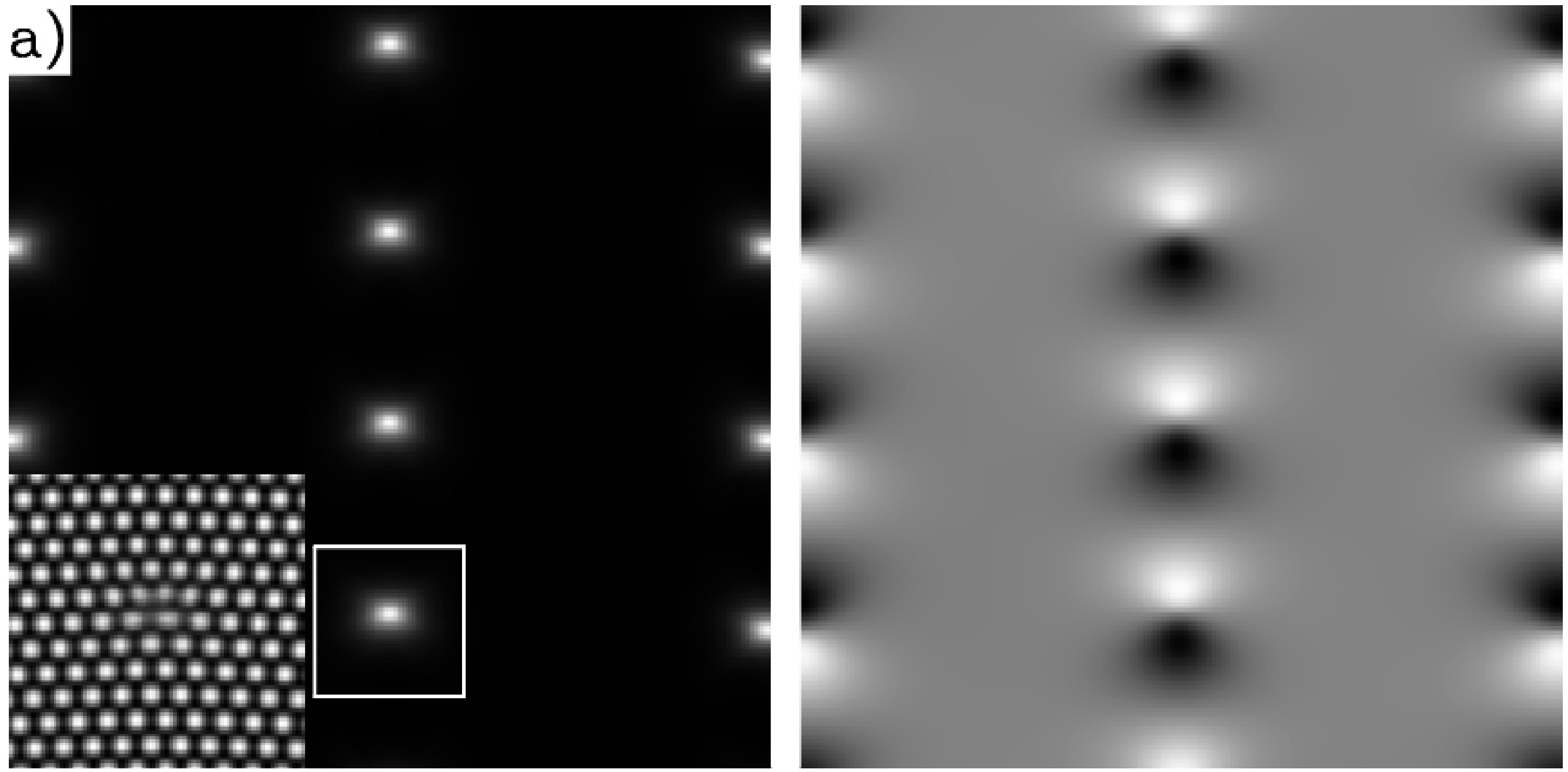}}
\centerline{\includegraphics[width=0.48\textwidth]{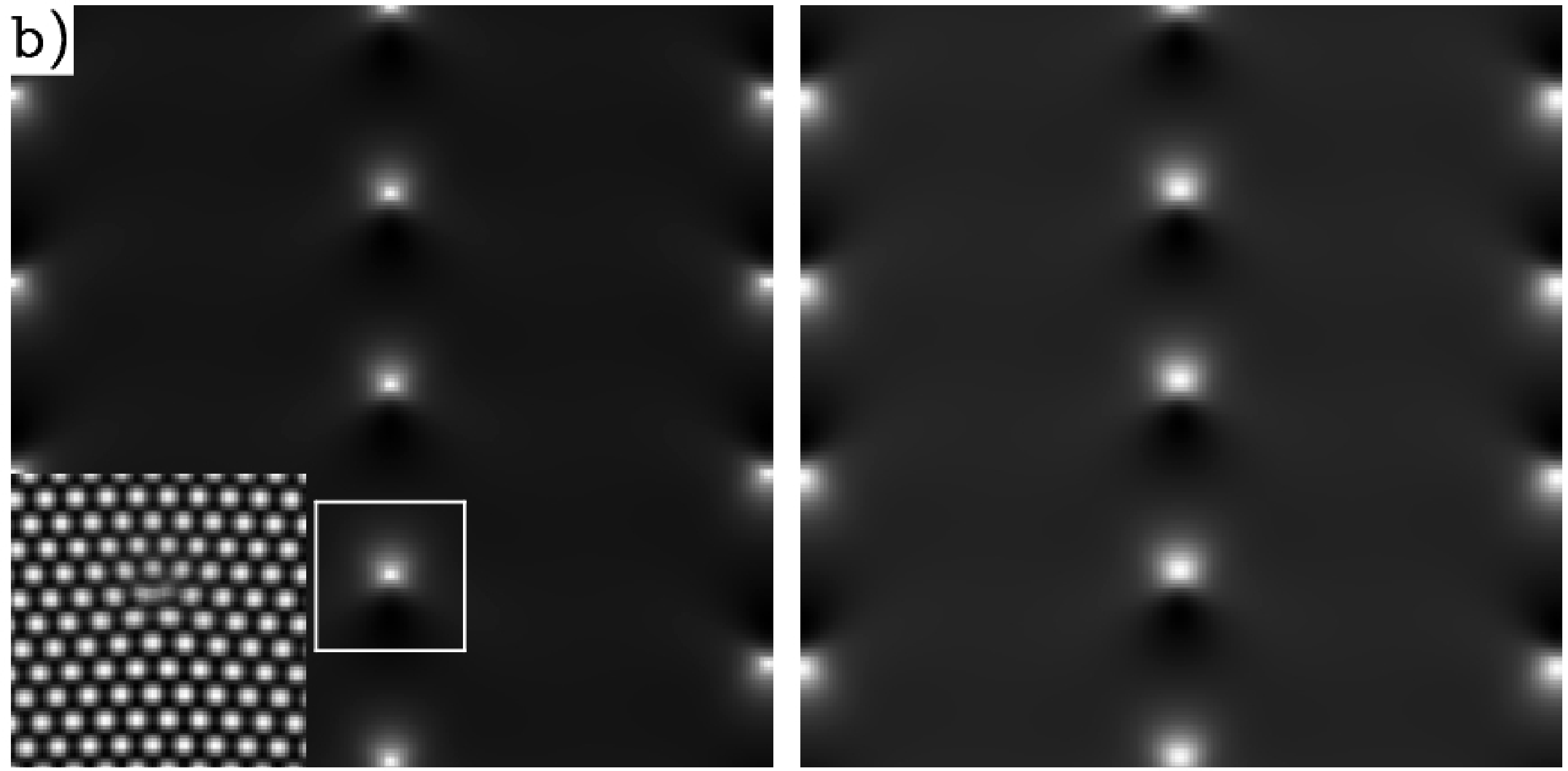}}
\caption{Solute segregation in symmetric grain boundaries with misorientation 
$\theta = 3.76^{\circ}$ are shown for ${\psi}=0$ (a) and 
${\psi}=0.2$ (b).  The left and right
panels correspond to $\sum_j |\eta_j|^2$ and $\psi$ respectively. 
In the corner insets of left panels the density field $n$ is reconstructed
from the amplitudes $\eta_j$ for the boxed region.
In the right panels the dark (light) color corresponds to 
the larger (smaller) of the atomic species.
The parameters for (a) and (b) correspond to Fig. \ref{fig:triph}a and 
Fig. \ref{fig:triph}b at $\Delta B_0=0.01$ respectively.}
\label{fig:bpfcgb}
\end{figure}

\begin{figure}[btp]
\centerline{\includegraphics[width=0.48\textwidth]{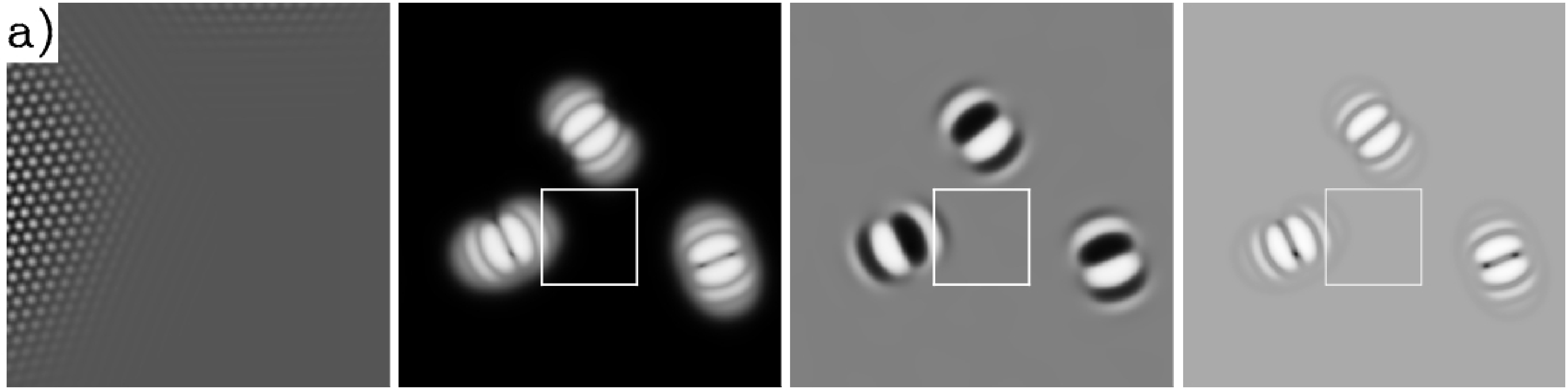}}
\centerline{\includegraphics[width=0.48\textwidth]{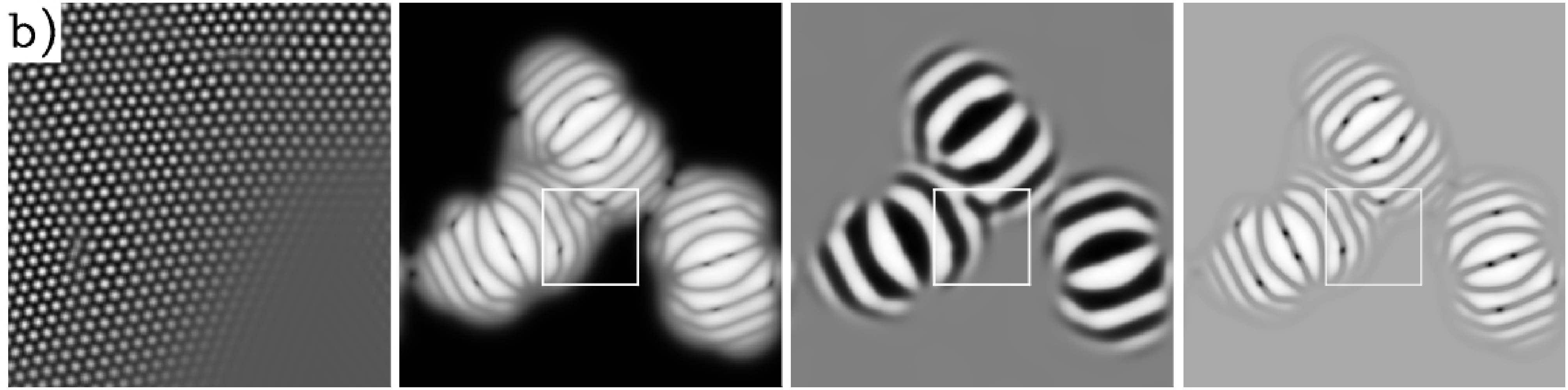}}
\centerline{\includegraphics[width=0.48\textwidth]{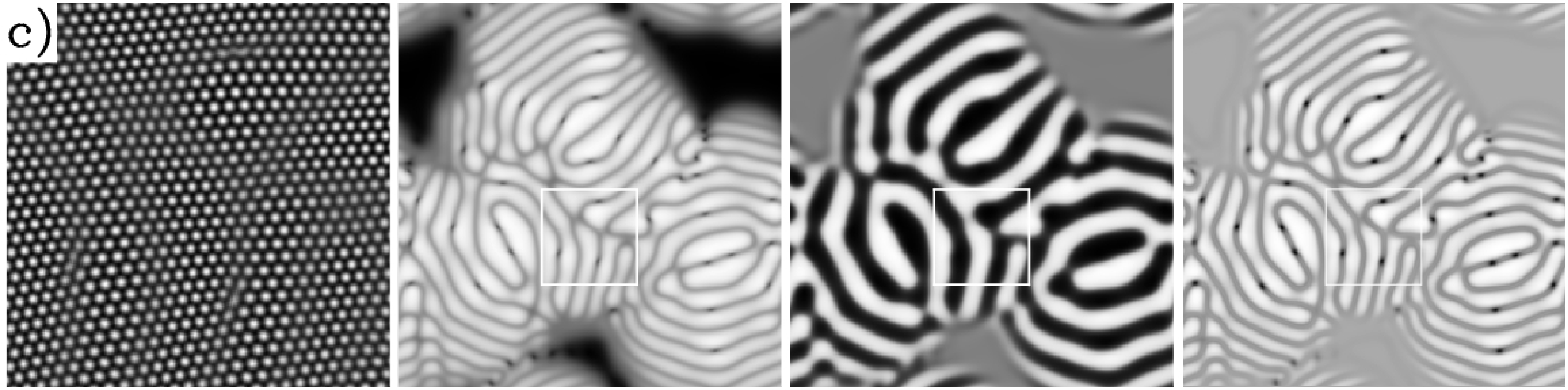}}
\caption{Eutectic solidification for parameters in Fig. \ref{fig:triph}b 
at $\Delta B_0=0.022$ and ${\psi} = 0.0$.  Panels (a), (b) and (c)
correspond to dimensionless times 30000, 60000 and 
105000 respectively.  From left to right the columns correspond to
$n$ reconstructed from $\eta_j$ (for the boxed region),
$\sum_j|\eta_j|^2$, $\psi$ and the local free energy density.  Dislocations 
are most easily identified as small black dots in the local free energy 
density.}
\label{fig:eutsol}
\end{figure}

\begin{figure}[btp]
\centerline{\includegraphics[width=0.48\textwidth]{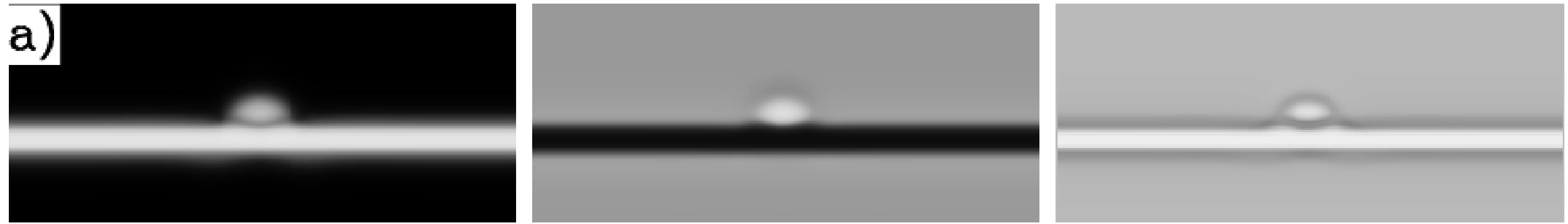}}
\centerline{\includegraphics[width=0.48\textwidth]{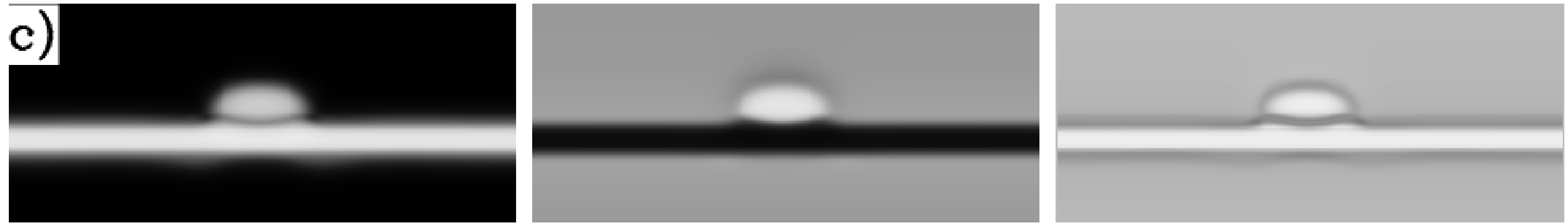}}
\centerline{\includegraphics[width=0.48\textwidth]{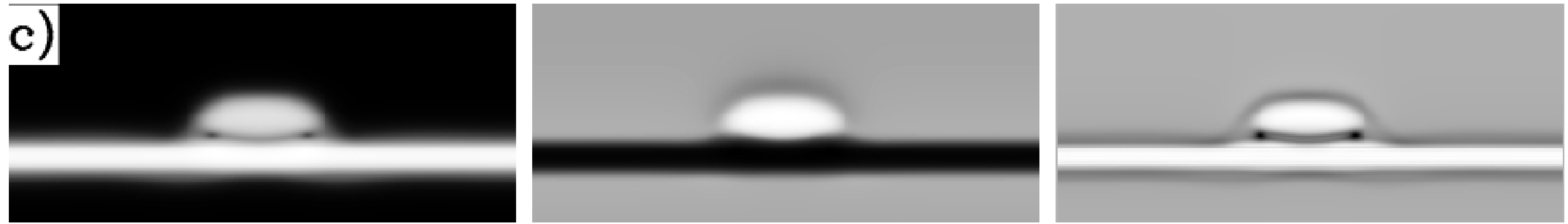}}
\caption{Quantum dot formation on a two atomic layer thick nanomembrane 
at times $t=20000$, $60000$ and $100000$ (for
(a) to (c) respectively). The columns from left to right 
correspond to $\sum|\eta_j|^2$, $\psi$ and the local free energy density.
The parameters used for this simulation 
are from Fig. \ref{fig:triph}a except $\alpha=0.26$ and 
$B^\ell_0 =  1.028$.}
\label{fig:qdot1}
\end{figure}

\begin{figure}[btp]
\centerline{\includegraphics[width=0.48\textwidth]{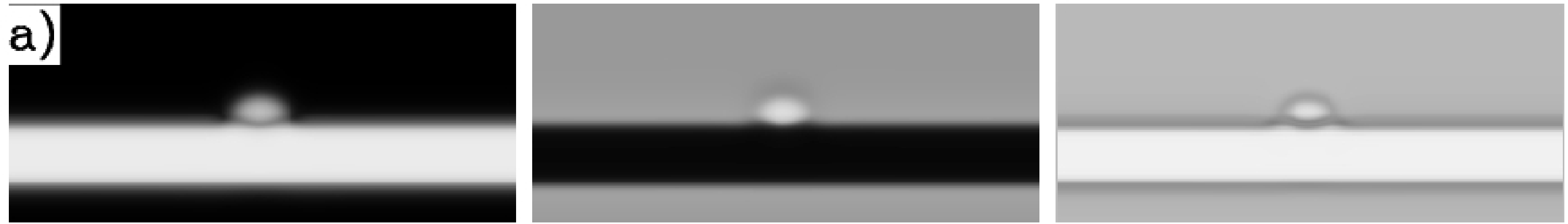}}
\centerline{\includegraphics[width=0.48\textwidth]{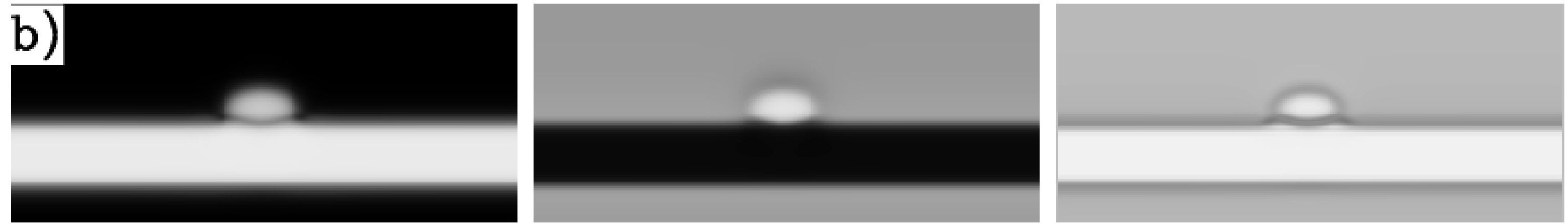}}
\centerline{\includegraphics[width=0.48\textwidth]{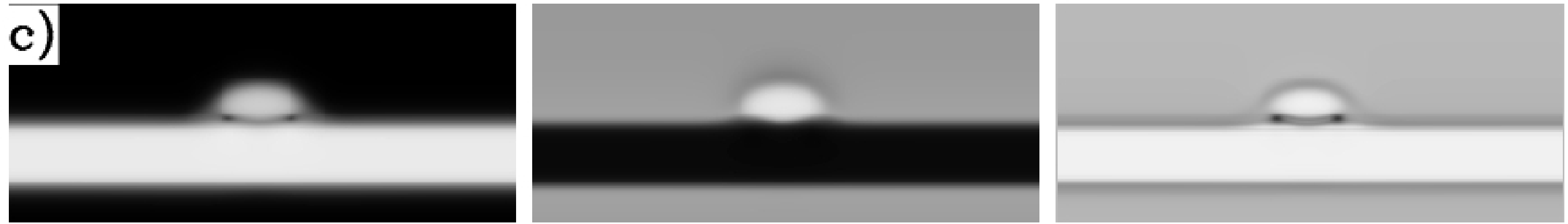}}
\caption{Quantum dot formation on a five atomic layer thick nanomembrane
at times $t=20000$, $40000$ and $60000$ (for 
(a) to (c) respectively). The parameters used 
are identical to those in Fig. \ref{fig:qdot1}.}
\label{fig:qdot2}
\end{figure}

\begin{figure}[btp]
\centerline{\includegraphics[width=0.48\textwidth]{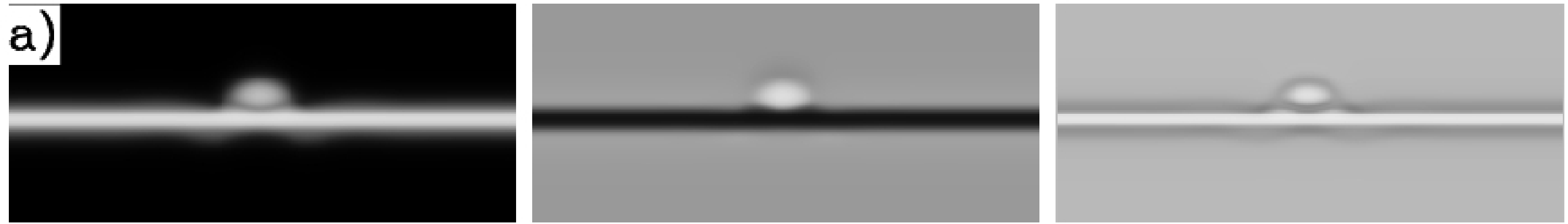}}
\centerline{\includegraphics[width=0.48\textwidth]{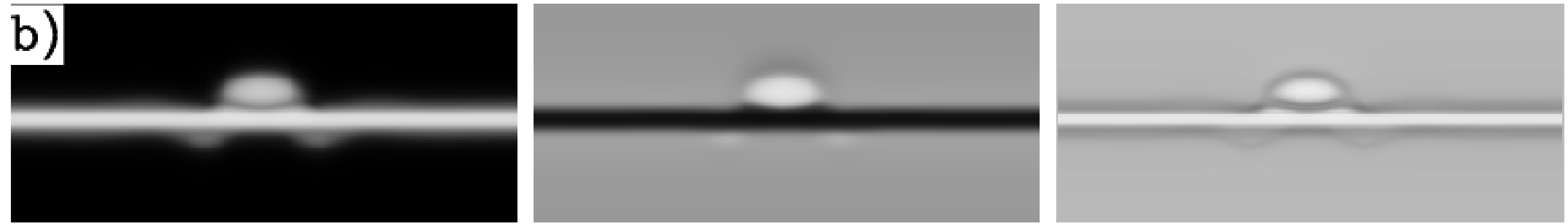}}
\centerline{\includegraphics[width=0.48\textwidth]{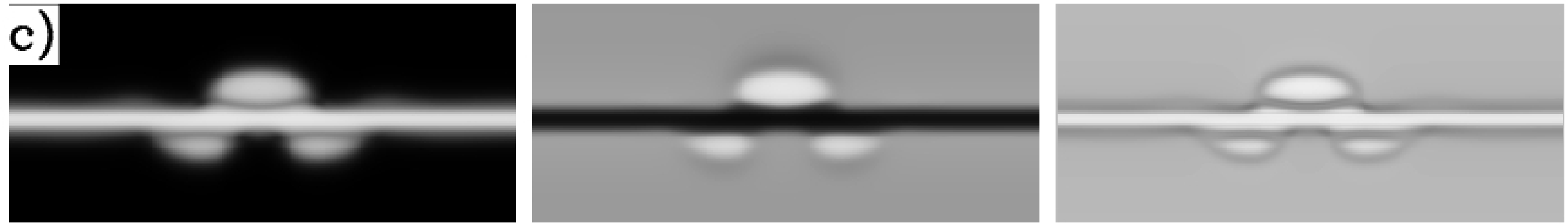}}
\centerline{\includegraphics[width=0.48\textwidth]{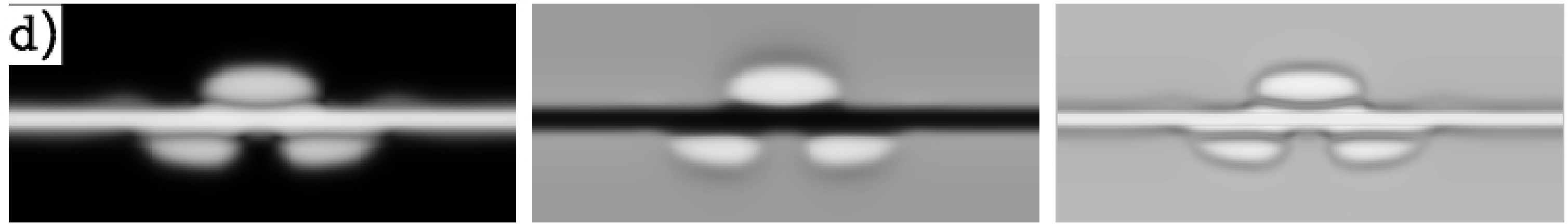}}
\centerline{\includegraphics[width=0.48\textwidth]{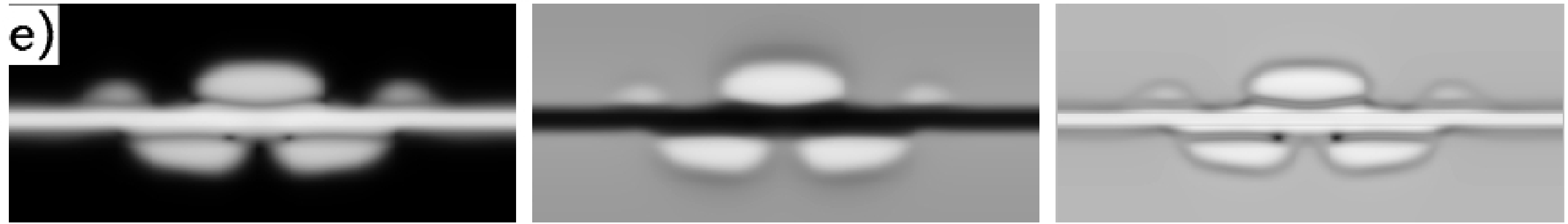}}
\centerline{\includegraphics[width=0.48\textwidth]{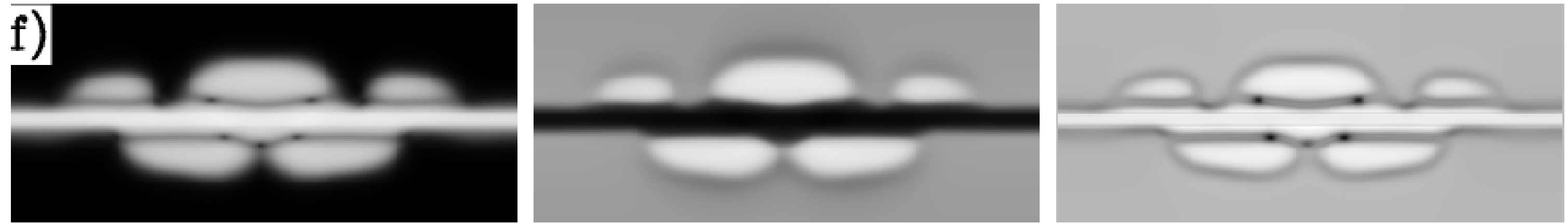}}
\caption{Correlated quantum dot formation on a nanomembrane at 
times $t=20000,40000,60000,80000,100000$ and $120000$ (for 
(a) to (f) respectively). The parameters used are identical 
to those in Fig. \ref{fig:qdot1} except for a higher liquid
supersaturation.}
\label{fig:multqdot}
\end{figure}

	To further examine the above amplitude equations,
numerical simulations of the model described 
by Eqs. (\ref{eq:modc}) and (\ref{eq:free_tri}) were undertaken and 
the results are shown in Figs. \ref{fig:bpfcgb}--\ref{fig:multqdot}.
In Fig. \ref{fig:bpfcgb} simulations were conducted to study the
coupling between composition and topological defects
(dislocations and grain boundaries) in binary alloys with components
of different atomic sizes, i.e., nonzero solute expansion coefficient
$\alpha$.  For this study a symmetric tilt grain boundary between 
two grains with a misorientation angle of $\theta=3.76^{\circ}$ was 
created by dynamically evolving an initial configuration of two perfect 
crystals separated by a layer of supercooled liquid.  As time evolves 
the liquid solidifies and a grain boundary spontaneously forms. 
The dislocation cores that comprise the grain boundary interact 
with the different atomic species or alloy composition. 
As shown in Fig. \ref{fig:bpfcgb} the
larger (smaller) solute atoms preferentially segregates 
around the dislocation cores in regions of tensile
(compressive) strain.

	Simulation results for eutectic solidification and phase
separation are presented in Fig. \ref{fig:eutsol}, where three small
grains of different orientation are heterogeneously nucleated for the
parameters used in Fig. \ref{fig:triph}b and for a ``temperature" $\Delta B_0$
below the eutectic point. As the grains grow the lamellar concentration 
bands form within the grains as a result of phase separation. 
The relatively large lattice mismatch (roughly $8.4\%$ in equilibrium,
due to a finite solute expansion coefficient) between the lamella
results in the spontaneous nucleation of dislocation at the lamellar
interfaces.  When the grains impinge on one another (i.e., coarsening
occurs) additional dislocations form, leading to complex patterns as
shown in Figs. \ref{fig:eutsol}b and \ref{fig:eutsol}c.
All these simulations indicate that the amplitude model established
here can simultaneously describe the complex evolution 
of liquid/solid interfaces, grain boundaries, dislocations and 
interfaces between regions of different compositions.

Numerical simulations were also conducted to study islands or 
quantum dots formation on thin freely standing films or solid nanomembranes. 
Recent experiments of Ge on Si membranes \cite{lag09a,lag09b} have
suggested that growth on such nanomembranes strongly influences the
maximum size that the strained islands can form coherently (i.e.,
without dislocations) and can lead to 
correlation or self-ordering of multiple islands.  To examine 
this phenomenon, simulations were set up such that islands of 
one material were grown
on a thin free standing membrane of another material, by exploiting a
eutectic phase diagram (such as that shown in Fig. \ref{fig:triph}b).  
To initiate growth of the islands a small crystal region at
${\psi}=-0.12$ was constructed coherently on top of a thin
membrane at ${\psi} = 0.2$ in a supercooled liquid 
at ${\psi}=-0.03$, with $\Delta B$ set to be $0.028$. 
Sample simulation results are shown in Figs. \ref{fig:qdot1}
and \ref{fig:qdot2}. As can be seen in these figures the dot grows 
coherently with the membrane until reaching some critical size at
which dislocations are nucleated at the liquid/dot/membrane junction.  
Perhaps more interestingly
by comparing Figs. \ref{fig:qdot1} and \ref{fig:qdot2} it is apparent 
that for the thinner membrane the dots can grow to a larger size
before nucleating dislocations.  This result occurs as the 
strained quantum dots partially relax by straining the substrate 
membrane and thinner membranes are easier to deform than thicker 
ones.
Such mechanism has been proved to play an important role in
engineering the self assembly of thin film
nanostructures such as quantum dots \cite{lag09a}.

	Another interesting consequence of the membrane deformation is
that it locally creates favorable and unfavorable positions for the
nucleation of other islands/dots, with an alternative sequence on the
two sides of the membrane, as can be seen in
Fig. \ref{fig:multqdot} which was conducted at a higher liquid
supersaturation (i.e., ${\psi}=-0.04$) or growth rate. 
Once a dot is formed, e.g., on the top of the membrane, it is
preferable for the next ones to nucleate under the edges rather than 
directly underneath the top quantum dot.  After these bottom-side dots 
are formed they in turn create preferential regions for a new 
set of dots to nucleate on the top (above their edges) and the process
repeats.  In such a fashion the process leads to the correlated dot growth 
as observed in experiments of Ge or SiGe islands 
on Si membranes \cite{lag09a,lag09b}. This phenomenon can 
in principle be exploited to create periodic strained nanostructures 
that in turn produce periodically modulated band gaps.

	Finally it should be noted that in the dynamics described by
Eq. (\ref{eq:modc}) the diffusion constant for the concentration field
($\psi$) is similar in magnitude in the liquid 
and solid phases.  However it is typically the case that diffusion 
or atomic mobility in liquids is much larger than in solids.  This
difference can be incorporated by using the following dynamic equation
for $\psi$, 
\be
\frac{\partial \psi}{\partial t} = \vec{\nabla} \cdot \left(  M(A) 
\vec{\nabla} \frac{\delta F}{\delta \psi}\right),
\ee
where the mobility now depends on $A^2=\sum_j|\eta_j|^2$.  
The following functional 
form,
\be
M = M_{X} + (M_{L}-M_{X})\left[1-\tanh\left(aA\right)\right]
\ee
would for example change $M$ from the mobility of the crystal 
($M_X$) at large $aA$ to the mobility in the liquid 
phase ($M_L$) at small or zero $aA$. The parameter 
$a$ would control how quickly the $M$ changes from liquid to 
solid phases.  In addition, while the binary PFC model presented 
in Eq. (\ref{eq:bpfc}) assumes both atomic species have the 
same mobility, the concentration dependence of $M$ could 
also be added in the above {\it ad hoc} method suggested for 
the phase dependence of atomic mobility.

\section{Discussions and Conclusions}

In this paper amplitude equations have been derived from the 
binary phase field crystal model for a two dimensional triangular 
lattice and for three dimensional BCC and FCC structures. Furthermore,
the connection to standard phase field models has been established,
and for small deformations the results have been shown to recover
linear continuum elasticity theory and reconstruct the equilibrium
phase diagrams for binary alloy systems. Sample simulations of the 
amplitude equations have shown
that this relatively simple model 
can effectively model many complex phenomena and the emergent microstructures
that arise, and reveal the underlying mechanisms.  While these 
amplitude equations were derived from an atomistic model, they can 
in themselves be regarded as phase field models with complex 
order parameters. 
One advantage  of this amplitude description is that the liquid and 
solid phases are easily distinguished by a relative uniform 
quantity $A^2\equiv \sum_j|\eta_j|^2$, and the coupling and interaction
between structure (i.e., amplitudes) and concentration can be well
identified.  In this regard it may be
possible to extend the equations to naturally incorporate other 
uniform fields, such as magnetization, polarization, temperature, etc. 
and simultaneously include elastic and plastic deformation in 
polycrystalline samples.

\section{Acknowledgements}
K.R.E. acknowledges support from NSF under Grant No. DMR-0906676.  
Z.-F.H. acknowledges support from NSF under Grant No. CAREER DMR-0845264.
N. P. acknowledges support from the National Science and 
Engineering Research Council of Canada

\appendix
\section{Amplitude Equations for BCC symmetry}
\label{sec:BCC}

For a BCC lattice, the principle reciprocal lattice vectors are
\ben
\vec{q}_1 &=& q_{eq}(\hat{x}+\hat{y}),  \nline
\vec{q}_2 &=& q_{eq}(\hat{x}+\hat{z}),  \nline
\vec{q}_3 &=& q_{eq}(\hat{y}+\hat{z}), 
\een
where $q_{eq} = 1/\sqrt{2}$ is the equilibrium wave number.  
In a one mode approximation the above principle reciprocal lattice
vectors need to be combined with the following vectors, 
$\vec{q}_4 = \vec{q}_1-\vec{q}_2 = q_{eq} (\hat{y}-\hat{z})$,
$\vec{q}_5 = \vec{q}_2-\vec{q}_3 = q_{eq} (\hat{x}-\hat{y})$ and
$\vec{q}_6 = \vec{q}_3-\vec{q}_1 = q_{eq} (-\hat{x}+\hat{z})$.  
Thus the BCC structure can be represented in the usual 
manner, i.e.,
\be
n = \sum_{j=1}^{j=6} \eta_j(\vec{r},t) e^{i\vec{q}_j\cdot\vec{r}} 
 + \sum_{j=1}^{j=6} \eta^*_j(\vec{r},t) e^{-i\vec{q}_j\cdot\vec{r}}. 
\ee
Repeating the calculations presented in section~(\ref{sec:amp}) 
(with the same level of approximations) gives the following complex 
amplitude equations:
\ben
\frac{\partial \eta_1}{\partial t} &=& 
-\left[\left(\Delta B_0+B^x_0{\cal G}^2_{1} 
+3v\left(\amp^2
-|\eta_1|^2\right)\right)\eta_1 
\right.\nline&&\left.
-2t(\eta_3\eta_6^*+\eta_2\eta_4)
+ 6v(\eta_3\eta_4\eta_5+\eta_2\eta_5^*\eta_6^*)
\right. \nline && \left. 
-2\alpha B^x_0\left(\psi  {\cal G}_1\eta_1
+{\cal G}_1\eta_1\psi \right)
\right],
\label{eq:bcca1}
\een
\ben
\frac{\partial \eta_4}{\partial t} &=& 
-\left[\left(\Delta B_0+B^x_0{\cal G}^2_{4} 
+3v\left(\amp^2-|\eta_4|^2\right)\right)\eta_4 
\right. \nline && \left.
-2t(\eta^*_5\eta_6^*+\eta_1\eta_2^*)
+ 6v(\eta_1\eta_3^*\eta_5^*+\eta_3\eta_2^*\eta_6^*)
\right. \nline && \left.
-2\alpha B^x_0\left(\psi {\cal G}_4 \eta_4
+{\cal G}_4\eta_4\psi \right)
\right],
\label{eq:bcca4}
\een
with equations of motion for $\eta_2$ and $\eta_3$ 
obtained by cyclic permutations on the groups (1,2,3) and (4,5,6)
in Eq. (\ref{eq:bcca1}), while equations for $\eta_5$ and 
$\eta_6$ can be obtained by the similar cyclic permutations 
of Eq. (\ref{eq:bcca4}). The corresponding concentration equation is given by
\ben
\frac{\partial \psi}{\partial t} 
&=&\nabla^2\left(
\left(w+B^\ell_2\amp^2 -K\nabla^2\right)\psi + u \psi^3  
\right.\nline && \left.
 -2B^x_0 \alpha\, 
\sum_j\left[
\eta_j{\cal G}^*_j\eta^*_j+
\eta^*_j{\cal G}_j\eta_j\right] 
\right). \nline
\label{BCCpsi}
\een

\begin{figure}[btp]
\includegraphics[width=0.5\textwidth]{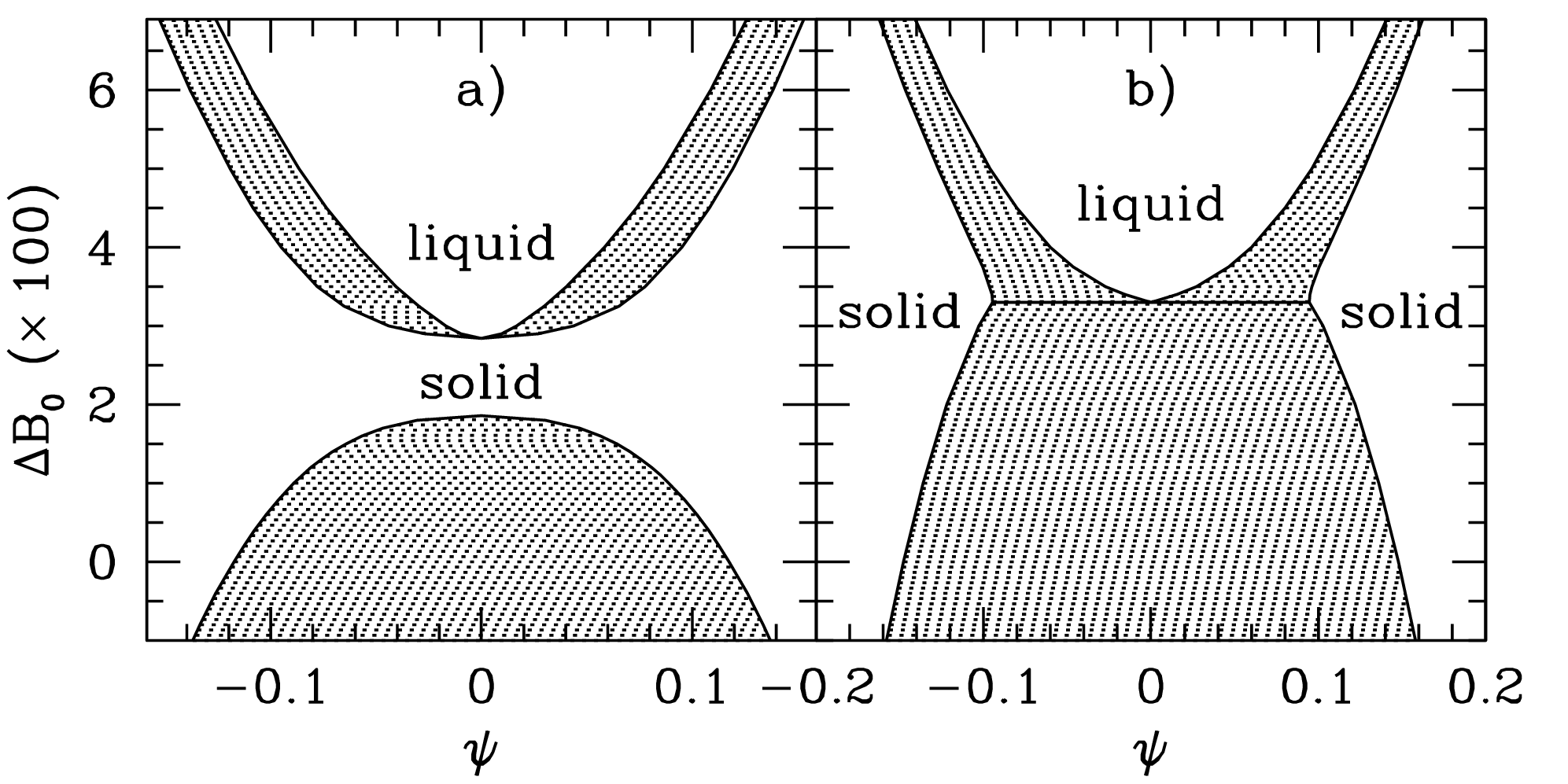}
\caption{Sample phase diagram for three dimensional 
BCC system with same parameter set and notation as in 
Fig. \ref{fig:triph} except $\omega=0.05$ in (a).  }
\label{fig:BCCph}
\end{figure}

Once again, Eqs.~(\ref{eq:bcca1})-(\ref{BCCpsi}) can 
be rewritten in a `model C' type form,
i.e.,
\ben
\frac{\partial \eta_j}{\partial t} 
= - \frac{\delta F}{\delta \eta_j^*}, \ \ \ \ \ \ 
\frac{\partial \psi }{\partial t} 
=  \nabla^2 \frac{\delta F}{\delta \psi},
\een
where
\ben
F &=& \int d\vec{r}\left[
\frac{\Delta B_0}{2}\amp^2+\frac{3v}{4}\amp^4
\right.\nline && \left.
+ \sum_{j=1}^{6}\left\{B^x_0|{\cal G}_j\eta_j|^2
-\frac{3v}{2}|\eta_j|^4\right\} 
\right.\nline && \left.
+6v\left(
\eta_1\eta_3^*\eta_4^*\eta_5^*
+\eta_2\eta_1^*\eta_5^*\eta_6^*
+\eta_3\eta_2^*\eta_6^*\eta_4^* + c.c.\right)
\right. \nline && \left.
-2t\left(\left[
\eta_1^*\eta_2\eta_4 + 
\eta_2^*\eta_3\eta_5 + 
\eta_3^*\eta_1\eta_6 + c.c.\right]
 \right.\right.\nline && \left.\left.
+\left[\eta^*_4\eta^*_5\eta^*_6+c.c.\right]\right)
\right.\nline && \left.
-2B^x_0\alpha\sum_{j=1}^6\left(
\eta_j {\cal G}^*_j\eta_j^*
+\eta^*_j {\cal G}_j\eta_j\right)\psi 
 \right.\nline && \left.
+\left(\omega + B_2^\ell\amp^2\right)\frac{\psi^2}{2} 
+\frac{u}{4}\psi^4 + \frac{K}{2}|\vec{\nabla}\psi|^2
\right].
\een

As discussed in Sec. \ref{sec:def} it is interesting to replace $\eta_j$ by 
$\phi\,e^{i\vec{q}_j\cdot\vec{r}}$, yielding
\ben
F &=& \int d\vec{r}\left[\left\{
6\Delta B_0\phi^2
-16t\phi^3
+135v\phi^4
\right.\right.\nline && \left.\left.
+\left(\omega + 12B_2^\ell\phi^2\right)\frac{\psi^2}{2} 
\right.\right.\nline&&\left.\left.
+\frac{u}{4}\psi^4 \right\}
-\left\{ \frac{K}{2}\psi\nabla^2\psi
+8B^x(1+3\alpha\psi)\phi\nabla^2\phi\right\}
\right. \nline && \left.
+4B^x_0\left\{
\sum_{i=1}^3 \left(U_{ii}^2+4\alpha\psi U_{ii}
+\frac{1}{2}\sum_{j\neq i}^3U_{ii}U_{jj}\right)
\right.\right. \nline && \left.\left.
+2\sum_{i=4}^6 U_{ii}^2
\right\}\phi^2 
\right].
\een
Similarly the elastic constants can be derived as
$C_{11}=C_{22}=C_{33}=8B^x_0\phi^2$ and 
$C_{12}=C_{13}=C_{23}= C_{44}=C_{55}=C_{66}=C_{11}/2$. 
Furthermore, minimizing with respect to $\delta$ gives,
$\delta_{eq}=-\alpha\psi$ leads to the following result for the free
energy per unit volume,
\ben
\frac{F}{\cal V} &=& 6\Delta B_0\phi^2 -16t\phi^3 +135v\phi^4 
\nline &&
+(\omega+6B^\ell_2\phi^2)\frac{\psi^2}{2}
\nline &&
+\frac{u}{4}\phi^4
-24B^x_0(\alpha\psi\phi)^2,
\een
which can be minimized with respect to $\phi$ to obtain
\be
\phi_{eq}=\frac{2t+\sqrt{4t^2-45v(\Delta B_0 
+\psi^2(B^\ell_2-4B_0^x\alpha^2)}}{45v}.
\ee
The corresponding phase diagram for this model is shown in
Fig. \ref{fig:BCCph}.

\section{Amplitude Equations for FCC symmetry}
\label{sec:FCC}

Recently Wu \cite{Wu06} has shown 
that the basic phase field crystal model can be
extended to model a FCC lattice by including 
an extra length scale.  Replacing 
the operator $\Delta B+B^x(1+R^2\nabla^2)^2 \rightarrow $ 
${\cal D} \equiv \Delta B+16\,B^x(1+2R_1^2\nabla^2+R_1^4\nabla^4)(
1+2R^2_2\nabla^2+R_2^4+E)$, 
where $R_2=\sqrt{3}/2 R_1$ for an FCC lattice and 
the extra factor of $16$ is introduced just for convenience,
gives a revised alloy PFC free energy,
\ben
F &=& \int d\vec{r} \left\{\frac{n{\cal D}n}{2}
-\frac{t}{3}n^3+\frac{v}{4}n^4
\right. \nline && \left. 
+\frac{\omega}{2}\psi^2 + \frac{u}{4}\psi^4 + \frac{K}{2}|\vec{\nabla}\psi|^2
\right\}.
\een
Here the parameter $E$ controls the symmetry of the phase such that 
for large (small) $E$ a BCC (FCC) structure is favored.  The
following calculations focus on the FCC phase in the limit of $E=0$.
Furthermore, to examine the influence of solute expansion the 
parameters $R_1$ and $R_2$ can be set to be $R_1=1+\alpha\psi$ 
and $R_2=\sqrt{3}/2(1+\alpha\psi)$. In the small $\alpha\psi$ 
limit the free energy functional becomes,
\ben
F &=& \int d\vec{r} \left\{\frac{n\genop{0}n}{2}
+\alpha \psi \frac{n\genop{1} n}{2}
-\frac{t}{3}n^3+\frac{v}{4}n^4
\right.\nline && \left.
+\frac{\omega}{2}\psi^2 
 + \frac{u}{4}\psi^4 + \frac{K}{2}|\vec{\nabla}\psi|^2
\right\},
\een
showing the same form as Eq. (\ref{eq:lazy1}) but with different operators
\ben
\genop{0} &\equiv& \Delta B_0 +
B^\ell_2\psi^2+16\,B^x(1+\nabla^2)^2(1+3/4\nabla^2)^2, \nline
\genop{1} &\equiv& 16\,B^x\nabla^2(1+\nabla^2)(1+3/4\nabla^2)(7+6\nabla^2).
\een
The corresponding equations of motion are also governed by
Eqs. (\ref{eq:ndyn}) and (\ref{eq:psidyn}), with $\genop{0}$ and
$\genop{1}$ given above.

For the FCC symmetry the principle reciprocal lattice vectors are
\ben
\vec{q}_1 &=& q_{eq}(-\hat{x}+\hat{y}+\hat{z})/\sqrt{3}, \nline
\vec{q}_2 &=& q_{eq}(\hat{x}-\hat{y}+\hat{z})/\sqrt{3}, \nline
\vec{q}_3 &=& q_{eq}(\hat{x}+\hat{y}-\hat{z})/\sqrt{3}. 
\een
To construct an FCC crystal the following reciprocal lattice vectors are also required,  
\ben
\vec{q}_4 &=& -\vec{q}_1-\vec{q}_2-\vec{q}_3 
= q_{eq}(-\hat{x}-\hat{y}-\hat{z})/\sqrt{3}, \nline
\vec{q}_5 &=& \vec{q}_1+\vec{q}_2 = 2q_{eq}\hat{z}/\sqrt{3},   \nline
\vec{q}_6 &=& \vec{q}_2+\vec{q}_3 = 2q_{eq}\hat{x}/\sqrt{3},   \nline
\vec{q}_7 &=& \vec{q}_3+\vec{q}_1 = 2q_{eq}\hat{y}/\sqrt{3} .  
\een
Unlike the triangular and BCC symmetries, the FCC lattice requires 
at minimum two set of vectors of different lengths, i.e.,
$(\vec{q}_1,\vec{q}_2,\vec{q}_3,\vec{q}_4)$ with length $q_{eq}$ and 
$(\vec{q}_5,\vec{q}_6,\vec{q}_7)$ with length $2/\sqrt{3}\,q_{eq}$.   
The density field $n$ is then expanded in the usual fashion, i.e., 
\be
n = \sum_{j=1}^{j=7} \eta_j(\vec{r},t) e^{i\vec{q}_j\cdot\vec{r}} 
+ {\rm c.c.}
\ee

Following the standard procedure, the amplitude equations can be
derived as
\ben
\frac{\partial \eta_1}{\partial t} &=& \caleta{1}
\left[
\genopj{0}{1}  
\eta_1 -2t 
\left(
\eta^*_2 \eta_5 
+\eta^*_3 \eta_7+
\eta^*_4 \eta^*_6
\right) 
\right.\nline&&\left.
+6v\left(
\left\{
A^2-|\eta_1|^2/2\right\}
\eta_1
+\eta^*_2\eta^*_3\eta^*_4
\right.\right.\nline&&\left.\left.
+\eta_2 \eta^*_6\eta_7
+\eta_3 \eta_5\eta^*_6
+\eta_4\eta_5\eta_7
\right) 
\right. \nline && \left.
+\frac{\alpha}{2}\left(\psi\genopj{1}{1}\eta_1+\genopj{1}{1}\eta_1\right) 
\right],
\een

\ben
\frac{\partial \eta_2}{\partial t} &=& \caleta{2}
\left[
\genopj{0}{2}\eta_2 -2t \left(
\eta^*_3 \eta_6 
+\eta^*_4 \eta^*_7+
\eta^*_1 {\eta}_5
\right) 
\right.\nline&&\left.
+6v\left(
\left\{
A^2-|\eta_2|^2/2\right\}
\eta_2
+\eta^*_1\eta^*_3\eta^*_4
\right.\right.\nline&&\left.\left.
+\eta_1 \eta_6 \eta^*_7 
+ \eta_3 \eta_5 \eta^*_7
+\eta_4\eta_5\eta_6
\right)
\right. \nline && \left.
+\frac{\alpha}{2}\left(\psi\genopj{1}{2}\eta_2+\genopj{1}{2}\eta_2\right) 
\right],
\een

\ben
\frac{\partial \eta_3}{\partial t} &=& \caleta{3}
\left[
\genopj{0}{3}\eta_3 -2t \left(
\eta^*_4 \eta^*_5 
+\eta^*_1 \eta_7+
\eta^*_2 \eta_6
\right) 
\right.\nline && \left.
+6v\left(
\left\{
A^2-|\eta_3|^2/2\right\}
\eta_3
+\eta^*_1\eta^*_2\eta^*_4
\right.\right.\nline&&\left.\left.
+\eta_1 \eta^*_5\eta_6
+\eta_2 \eta^*_5{\eta}_7
+\eta_4\eta_6\eta_7
\right)
\right. \nline && \left.
+\frac{\alpha}{2}\left(\psi\genopj{1}{3}\eta_3+\genopj{1}{3}\eta_3\right) 
\right],
\een

\ben
\frac{\partial \eta_4}{\partial t} &=& \caleta{4}
\left[
\genopj{0}{4}\eta_4 -2t \left(
\eta^*_1 \eta^*_6 
+\eta^*_2 \eta^*_7+
\eta^*_3 \eta^*_5
\right) 
\right.\nline && \left.
+6v\left(
\left\{
A^2-|\eta_4|^2/2\right\}
\eta_4
+\eta^*_1\eta^*_2\eta^*_3
\right.\right.\nline&&\left.\left.
+\eta_1 \eta^*_5\eta^*_7
+\eta_2 \eta^*_5\eta^*_6
+\eta_3\eta^*_6\eta^*_7
\right)
\right. \nline && \left.
+\frac{\alpha}{2}\left(\psi\genopj{1}{4}\eta_4+\genopj{1}{4}\eta_4\right) 
\right],
\een

\ben
\frac{\partial \eta_5}{\partial t} &=& \calnu{5}
\left[
\genopj{0}{5}\eta_5 -2t \left(
\eta_1 \eta_2 
+\eta^*_3 \eta^*_4 
\right) 
\right.\nline && \left.
+6v\left(
\left\{
A^2-|\eta_5|^2/2\right\}
\eta_5
+\eta_6\eta_1\eta^*_3
\right.\right.\nline&&\left.\left.
+\eta^*_6\eta_2\eta^*_4
+\eta_7\eta_2\eta^*_3
+\eta^*_7\eta_1\eta^*_4
\right)
\right. \nline && \left.
+\frac{\alpha}{2}\left(\psi\genopj{1}{5}\eta_5+\genopj{1}{5}\eta_5\right) 
\right],
\een

\begin{figure}[btp]
\includegraphics[width=0.5\textwidth]{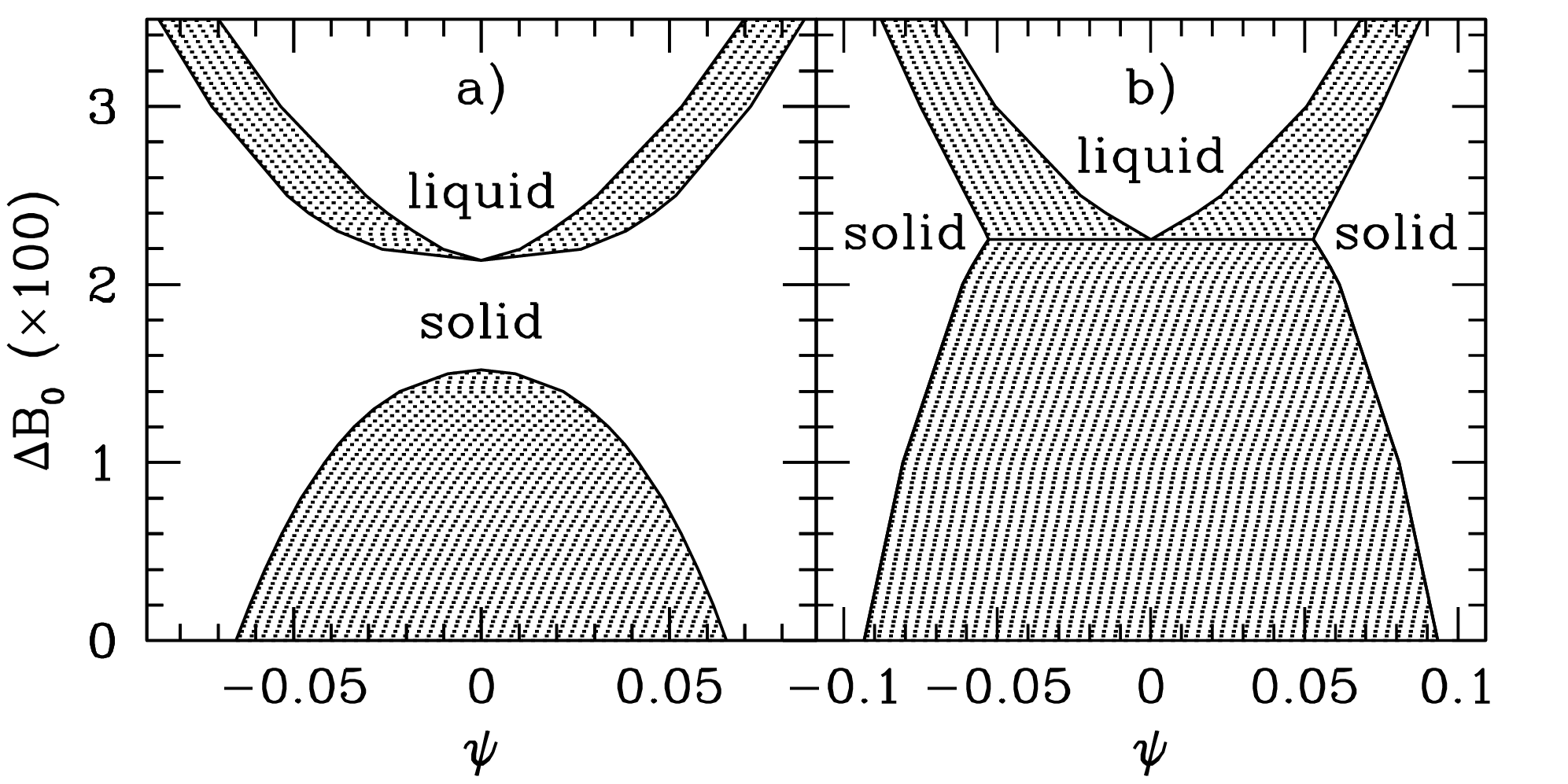}
\caption{Sample phase diagram for three dimensional 
FCC system with the same parameter set and notation as in 
Fig. \ref{fig:triph} except $\omega=0.02$ in (a).}
\label{fig:FCCph}
\end{figure}

\ben
\frac{\partial \eta_6}{\partial t} &=& \calnu{6}
\left[
\genopj{0}{6}\eta_6 -2t \left(
\eta_2 \eta_3 
+\eta^*_4 \eta^*_1 
\right) 
\right.\nline && \left.
+6v\left(
\left\{
A^2-|\eta_6|^2/2\right\}
\eta_6
+\eta_7\eta_2\eta^*_1
\right.\right.\nline&&\left.\left.
+\eta^*_7\eta_3\eta^*_4
+\eta_5\eta_3\eta^*_1
+\eta^*_5\eta_2\eta^*_4
\right)
\right. \nline && \left.
+\frac{\alpha}{2}\left(\psi\genopj{1}{6}\eta_6+\genopj{1}{6}\eta_6\right) 
\right],
\een

\ben
\frac{\partial \eta_7}{\partial t} &=& \calnu{7}
\left[
\genopj{0}{7}\eta_7 -2t \left(
\eta_1 \eta_3 
+\eta^*_2 \eta^*_4 
\right) 
\right. \nline && \left.
+6v\left(
\left\{
A^2-|\eta_7|^2/2
\right\}
\eta_7
+\eta_5\eta^*_2\eta_3
\right.\right.\nline&&\left.\left.
+\eta^*_5\eta_1\eta^*_4
+\eta_6\eta_1\eta^*_2
+\eta^*_6\eta_3\eta^*_4
\right)
\right. \nline && \left.
+\frac{\alpha}{2}\left(\psi\genopj{1}{7}\eta_7+\genopj{1}{7}\eta_7\right) 
\right],
\een

\ben
\frac{\partial \psi }{\partial t} 
&=&\nabla^2\left(
(w+B^\ell_2A^2-K\nabla^2)\psi
+ u \psi^3 
\right.\nline&&\left.
+\frac{\alpha}{2}\sum_j\left[
\eta_j(\genopj{1}{j})^*\eta_j^*+\eta_j^*\genopj{1}{j}\eta_j
\right] 
\right),
\een
where $\caleta{j}\equiv \nabla^2 + 2i\vec{q}_j\cdot\vec{\nabla} -|\vec{q}_j|^2$ , 
$A^2 \equiv 2\sum_{j} |\eta_j|^2$, and
\ben
\genopj{0}{j} &\equiv&
\Delta B_0 + B^\ell_2\psi^2+16B^x(1+{\cal L}_j)^2(1+3{\cal L}_j/4)^2,  \nline
\genopj{1}{j} &\equiv&
16B^x {\cal L}_j(1+{\cal L}_j)(1+3/4{\cal L}_j)(7+6{\cal L}_j).
\een
Note that $|\vec{q}_j|^2=1$ for $j=1,2,3,4$ 
and $|\vec{q}_j|^2=4/3$ for $j=5,6,7$, and thus we have
\ben
\genopj{0}{j} &=& \left\{
\begin{array}{cc}
 \Delta B_0 + B^\ell_2\psi^2+ B^x ({\cal G}_{j})^2(
3{\cal G}_j+1)^2 
& j=1,2,3,4 \\
 \Delta B_0+  B^\ell_2\psi^2+ B^x (3{\cal G}_j-1)^2
({\cal G}_{j})^2  
& j=5,6,7
\end{array}
\right.
\nline 
&\approx& \Delta B_0 + B^\ell_2\psi^2+ B^x ({\cal G}_{j})^2  \ \ \ j=1,...,7
\een

\ben
\genopj{1}{j} &=& \left\{
\begin{array}{cl}
 4^x ({\cal G}_{j}-1){\cal G}_j(3{\cal G}_j+1)(6{\cal G}_j+1)
& j=1,2,3,4 
\\
 4B^x ({\cal G}_{j}-4/3)(3{\cal G}_j-1){\cal G}_j(6{\cal G}_j-1)
& j=5,6,7
\end{array}
\right.
\nline
&\approx&
\left \{ \begin{array}{cl}
 -4B^x ({\cal G}_{j})
& j=1,2,3,4 
\\
 -4B^x (4{\cal G}_{j}/3)
& j=5,6,7
\end{array}
\right.
\een
where ${\cal G}_{j} \equiv \nabla^2+2i\vec{q}_j\cdot\vec{\nabla}$. 
Finally as described in Sec. \ref{sec:def} the operator $\calnu{j}$
acting on the right hand side of each equation can be approximated by
constants, i.e., $\caleta{j} \approx -|\vec{q}_j|^2$. 
As before the dynamical equations can be written as 
\ben
\frac{\partial \eta_j}{\partial t} &=& 
-\,|\vec{q}_j|^2\,\frac{\delta F}{\delta \eta_j^*}, \nline
\frac{\partial \psi}{\partial t} &=& 
\nabla^2\frac{\delta F}{\delta \psi},
\een
where
\ben
F &=& \int d\vec{r} 
\left[\frac{\Delta B_0}{2}\amp^2+\frac{3v}{2}\amp^4
\right.\nline && \left.
+\sum_{j=1}^{7}
\left\{B^x|{\cal G}_{j}\eta_j|^2
-\frac{9v}{2}|\eta_j|^4
\right\}
\right. \nline && \left. 
-2t
\left(
\eta^*_1(\eta^*_2\eta_5+\eta^*_3\eta_7+\eta^*_4\eta^*_6) 
+\eta^*_2(\eta^*_3\eta_6+\eta^*_4\eta^*_7)
\right.\right. \nline && \left. \left.
+\eta^*_3 \eta^*_4 \eta^*_5
+{\rm c.c.}
\right)
+6v \left(
\eta^*_1(
\eta^*_2 \eta^*_3 \eta^*_4
+\eta_2 \eta^*_6 \eta_7
\right.\right.  \nline && \left.\left.
+\eta_3 \eta_5 \eta^*_6
+\eta_4 \eta_5 \eta_7)
+\eta^*_2\eta_5(
\eta_3  \eta^*_7
+\eta_4  \eta_6)
\right.\right.  \nline && \left.\left.
+\eta^*_3\eta_4 \eta_6 \eta_7 + {\rm c.c.}
\right)
\right. \nline && \left. 
-2\alpha B^x_0 \left(
\sum_{j=1}^4 \eta^*_j {\cal G}_j \eta_j+\frac{4}{3}\sum_{j=5}^7 
\eta^*_j {\cal G}_j \eta_j + {\rm c.c.}\right)\psi
\right. \nline && \left. 
+(\omega+B^\ell_2 \amp^2)\frac{\psi^2}{2} + \frac{u}{4}\psi^4 
+ \frac{K}{2}|\vec{\nabla}\psi|^2
\right].
\een

Setting $\eta_j=\phi e^{i\vec{q}_j\cdot\vec{r}}$ as before gives
\ben
F &=& \int d\vec{r} 
\left[\left\{
7\Delta B_0 \phi^2
-24t\phi^3
+\frac{693v}{2}\phi^4
\right. \right.\nline &&\left.\left.
+(\omega+14B^\ell_2 \phi^2)\frac{\psi^2}{2} 
+ \frac{u}{4}\psi^4 \right\}
 \right.\nline &&\left.
-\left\{ \frac{K}{2}\psi\nabla^2\psi
+32B^x_0\left(\frac{1}{3}+\alpha\psi\right)\phi\nabla^2\phi\right\}
\right. \nline && \left. 
+\frac{16}{9}B^x_0\left\{
\sum_{i=1}^3\left(5U_{ii}^2+14\alpha\psi U_{ii}
+\sum_{j\neq i}^3 U_{ii}U_{jj}\right)
\right.\right.\nline&&\left.\left.
+4\sum_{i=4}^6U_{ii}^2
\right\}\phi^2
\right],
\een
from which the elastic constants for a FCC lattice can be obtained: 
$C_{11}=C_{22}=C_{33}=160B^x_0\phi^2/9$ and
$C_{12}=C_{13}=C_{23}=C_{44}=C_{55}=C_{66}=32 B^x_0\phi^2/9=C_{11}/5$.

If minimizing the above free energy expression with respect to
$\delta$, we obtain $\delta_{eq}=-\alpha\psi$ and hence
the following free energy per unit volume
\ben
\frac{F}{\cal V} &=& 7\Delta B_0 \phi^2 - 24t\phi^3+\frac{693}{2}v\phi^4
+(\omega+7B^\ell_2\phi^2)\frac{\psi^2}{2}
\nline 
&&+\frac{u}{4}\psi^4-\frac{112}{3}
B^x_0(\alpha\phi\psi)^2,
\een
which is minimized when
\be
\phi_{eq}=\frac{18t+\sqrt{324t^2-4851v(\Delta B_0
    +\psi^2(B^\ell_2-16B^x_0\alpha^2/3))}}{693v}.
\ee
A phase diagram for this FCC amplitude model is presented in
Fig. \ref{fig:FCCph}.

\end{document}